\documentstyle[preprint,aps]{revtex}

\begin{document}
\title{Two-waves approximation for reflected and transmitted energy in photonic
crystals}
\author{B. Guizal, D. Felbacq, R. Sma\^{a}li}
\address{LASMEA UMR-CNRS 6602\\
Complexe des C\'{e}zeaux\\
63177 Aubi\`{e}re Cedex, France}
\date{\today }
\maketitle

\begin{abstract}
We study electromagnetic waves scattering by a 2D photonic crystal made of a
stack of diffraction gratings. In case where there are only two propagative
modes in the crystal, we derive an explicit expression for the superior
(resp. inferior) envelope of the reflected (resp. transmitted) energy.
\end{abstract}

\noindent \tightenlines Photonic crystals are periodically modulated devices
having photonic band gaps \cite{joan,weisbuch}. The theoretical study of
these structures may be performed by use of Bloch waves theory,
multi-scattering theory \cite{moi,cent}\ or they can be modelled as stacks
of gratings \cite{maystre,mcphed}, which are the elementary layers of the
device (cf. fig. 1). We use this last approach in this note.

\noindent Two situations can be roughly distinguished in the scattering
behaviour of periodic structures: when the wavelength is of the same order
of the period, the resonant domain, there are infinitely many evanescent
modes and finitely many propagative modes in the structure (the number of
which being given by Bloch waves theory). As the wavelength increases, the
number of propagative modes diminishes and is finally reduced to $2$. In the
very large wavelength regime, the structure behaves as if it were
homogeneous. In case of stacks of gratings, that leads to replace one
grating by a homogeneous slab (fig.1) with some effective permittivity
depending on the polarization\cite{cent,moibou}. However, there is an
intermediate zone of wavelengths for which there are only two propagative
modes inside the gratings, but the wavelength is not large enough to use a
homogenization scheme (for instance $\lambda $ is only two or three times
the period). This work deals with this last situation. The crystal is made
as a stack of $N$ gratings (fig 1) with vacuum above and below.

\noindent We consider first an elementary layer. As there are only two
propagative waves, there exists a $2\times 2$ matrix ${\bf T}_{\lambda
,\theta }$ such that ${\bf T}\left( 
\begin{array}{c}
A^{+} \\ 
A^{-}
\end{array}
\right) =\left( 
\begin{array}{c}
B^{+} \\ 
B^{-}
\end{array}
\right) $ (see fig. 2 for the notations), the elements of ${\bf T}$ are
denoted $\left( {\bf t}_{ij}\right) $. This matrix can be computed
rigorously from a standard theory of diffraction by gratings (for instance
the Coupled Waves Method, see \cite{cwm} for example). The main property of
matrix ${\bf T}$ is that $\det \left( {\bf T}\right) =1$ \cite{wave} so that
if $\mu $ is an eigenvalue then $\mu ^{-1}$ is the other eigenvalue. We have
shown in another article \cite{fresnel} that in a layer characterized by a $%
2\times 2$ transfer matrix, the reflected and transmitted coefficients for $%
N $ layers can be obtained in close form : 
\begin{equation}
\begin{array}{l}
r_{N}\left( \lambda ,\theta \right) =f%
{\displaystyle{\mu ^{2N}-1 \over \mu ^{2N}-fg^{-1}}}%
\\ 
t_{N}\left( \lambda ,\theta \right) =\mu ^{N}%
{\displaystyle{\left( 1-fg^{-1}\right)  \over \mu ^{2N}-fg^{-1}}}%
\end{array}
\label{rt}
\end{equation}
\newline
where $\mu $ is an eigenvalue of ${\bf T}$ associated with an eigenvector $%
\left( u_{1},u_{2}\right) $, an eigenvector associated to $\mu ^{-1}$ is
denoted by $\left( v_{1},v_{2}\right) $ and 
\begin{equation}
\begin{array}{l}
f\left( \lambda ,\theta \right) =%
{\displaystyle{i\beta _{0}v_{1}-v_{2} \over i\beta _{0}v_{1}+v_{2}}}%
,\text{ }g\left( \lambda ,\theta \right) =%
{\displaystyle{i\beta _{0}u_{1}-u_{2} \over i\beta _{0}u_{1}+u_{2}}}%
\\ 
\text{and }\beta _{0}=%
{\displaystyle{2\pi  \over \lambda }}%
\cos (\theta )
\end{array}
\end{equation}
\newline
{\bf Remarks}:

1) It is worth noticing that functions $f$ and $g$ do not depend on a
particular choice of an eigenvector.

2) The expressions (\ref{rt}) are valid whatever the angle of incidence and
polarization, it is only the particular expression of matrix ${\bf T}$ that
changes.

\bigskip

\noindent The expressions given in (\ref{rt}) show that {\bf it suffices to
compute the transfer matrix of one layer to compute the transmitted and
reflected fields}. However, the reflected and transmitted energies are
highly oscillating functions which can be better represented by their
envelope (superior and inferior limits when $N$ tends to infinity). We have
shown in \cite{jmp} that the superior envelope $R_{\infty }$ of the
reflected energy, and conversely the inferior envelope of the transmitted
energy $T_{\infty }$ are given by : 
\begin{equation}
\begin{array}{l}
T_{\infty }=%
{\displaystyle{4-tr\left( {\bf T}\right) ^{2} \over \left( {\bf t}_{12}\beta _{0}-{\bf t}_{21}\beta _{0}^{-1}\right) ^{2}}}%
\\ 
R_{\infty }=1-T_{\infty }
\end{array}
\label{epi}
\end{equation}
\newline
We now turn to some numerical examples. We use as an elementary layer a
lamellar dielectric grating with an air layer below (see fig. 1). The
parameters are the following: $h/\lambda =1$, $\varepsilon _{a}=4$ and $%
\varepsilon _{b}=1$ with a filling factor of $1/2$. It is illuminated by an
s-polarized monochromatic plane wave under normal incidence. We have plotted
in fig. 3 the reflected energy versus $\lambda /d,$ obtained for a stack of $%
20$ layers. We see that above $\lambda _{c}/d=1.515$ the true reflection
ratio is well approximated by (\ref{rt}): this corresponds to the critical
wavelength above which there are only two modes in the structure (cf. fig. 4
for the band structure). Above $\lambda _{c}/d$\ we see an almost perfect
agreement between the reflection computed from (\ref{rt}) and the reflection
obtained from a rigorous code (fig. 3). Finally, we have plotted the
envelope of the reflected energy obtained from (\ref{epi}), we see a perfect
agreement above wavelength $\lambda _{c}/d$.

\noindent We have developed an approximate theory for the study of photonic
band gap materials in case where there are two propagative modes in the
structure in both $s$- and $p$-polarization. This method allows explicit
computations of the field once the $2\times 2$ transfer matrix of one layer
is known. This last matrix can be obtained numerically, then equations (\ref
{rt}) and (\ref{epi}) can be used to obtain close-form expressions of the
diffracted field. This can be useful for studying the so-called
''ultra-refractive'' effects\cite{ultra,gralak,enoch}.

\newpage

\begin{center}
{\bf Figure captions}
\end{center}

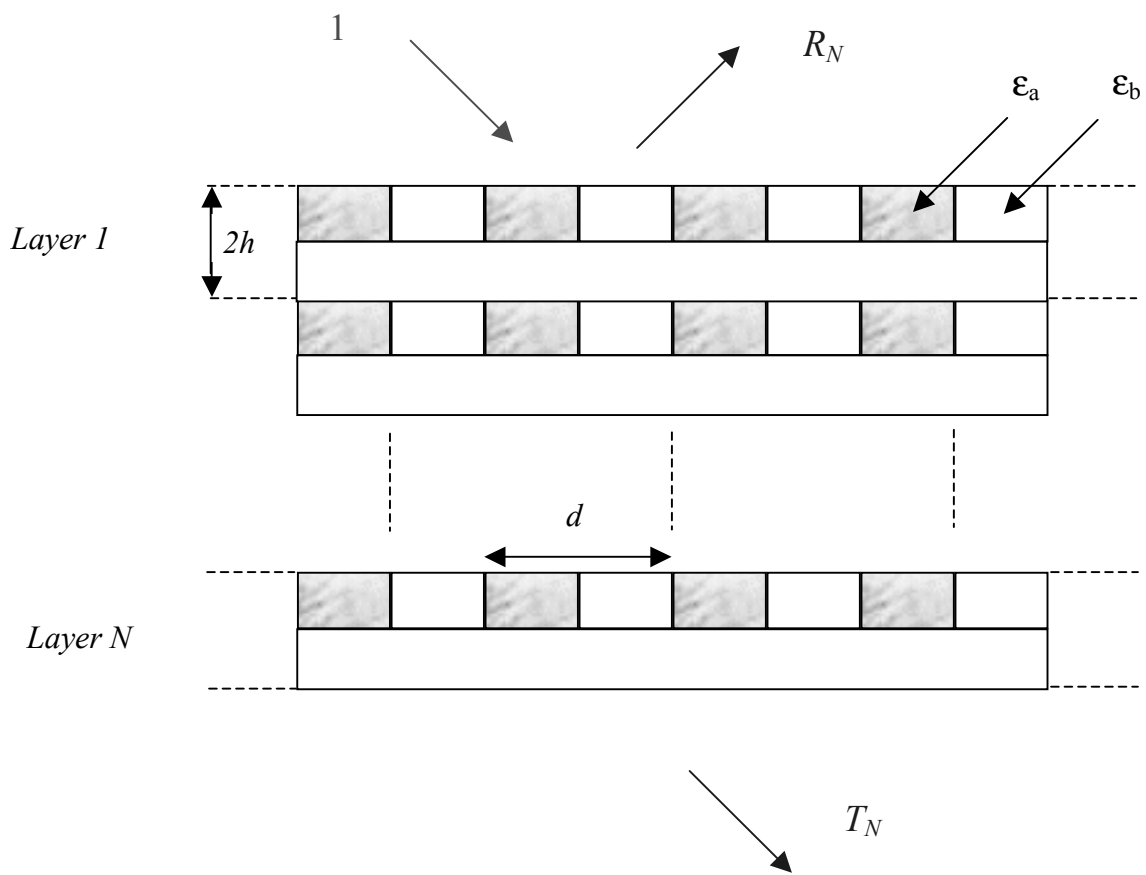
{\bf Figure 1}: Geometry of the photonic structure : the elementary cell is
made of a grating, with period $d$ filled with materials whose relative
permittivities are $\varepsilon _{a}$ and $\varepsilon _{b}$. There is a
dielectric layer below.

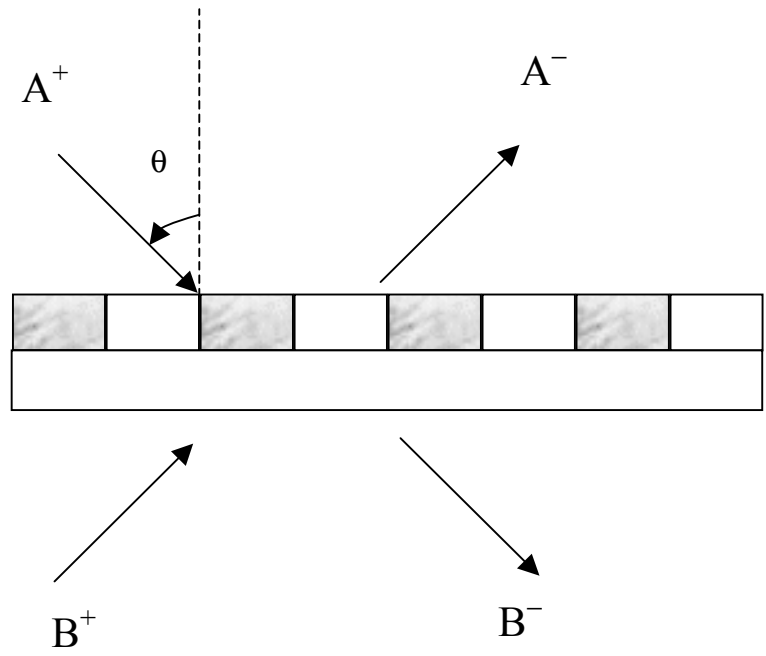
{\bf Figure 2}: Definition of the transfer matrix of a layer.

{\bf Figure 3}: Reflection vs $\lambda /d$. Solid lines : reflection
obtained from a rigorous theory.\ Dots : reflection obtained from formulas (%
\ref{rt}).

{\bf Figure 4}: Band structure of the photonic device: when $\lambda
>\lambda _{c}$ there are only two propagative waves in the crystal.

{\bf Figure 5}: (a) Envelope of the reflected energy (bold line) obtained
from (\ref{epi})\ and reflected energy (solid line).

\qquad \qquad\ (b) Zoom of figure 5 (a) around the bifurcation value $%
\lambda _{c}/d$.

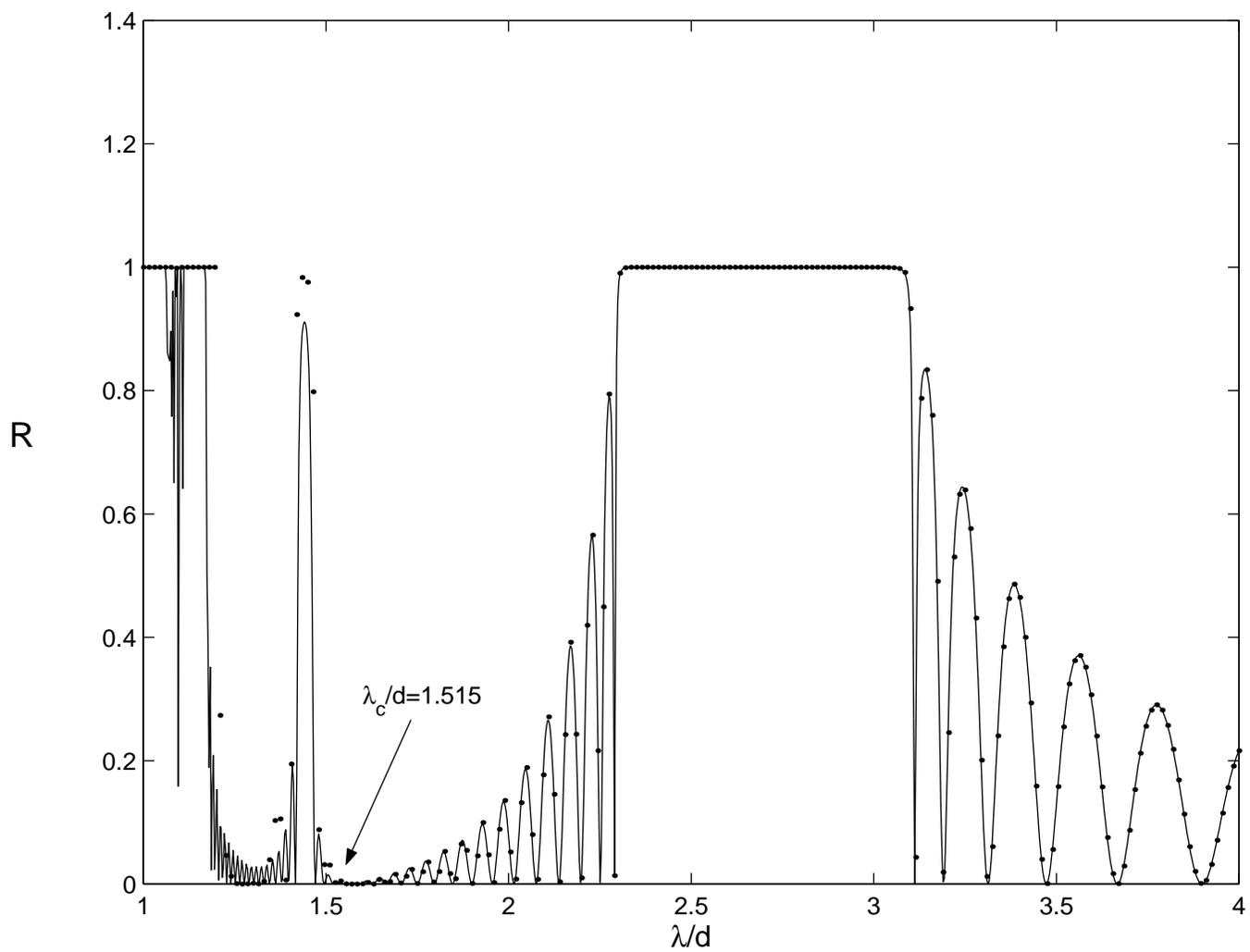
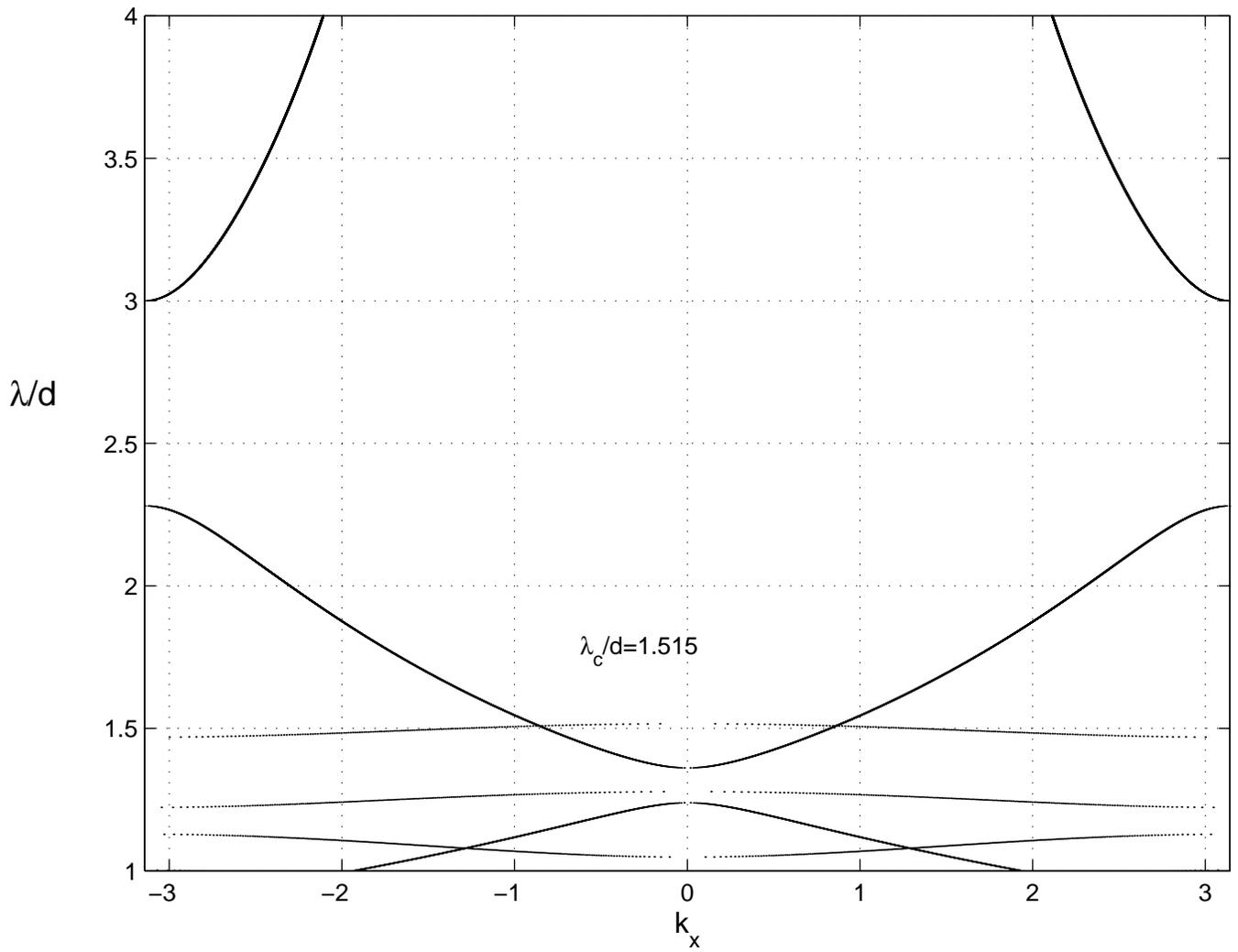
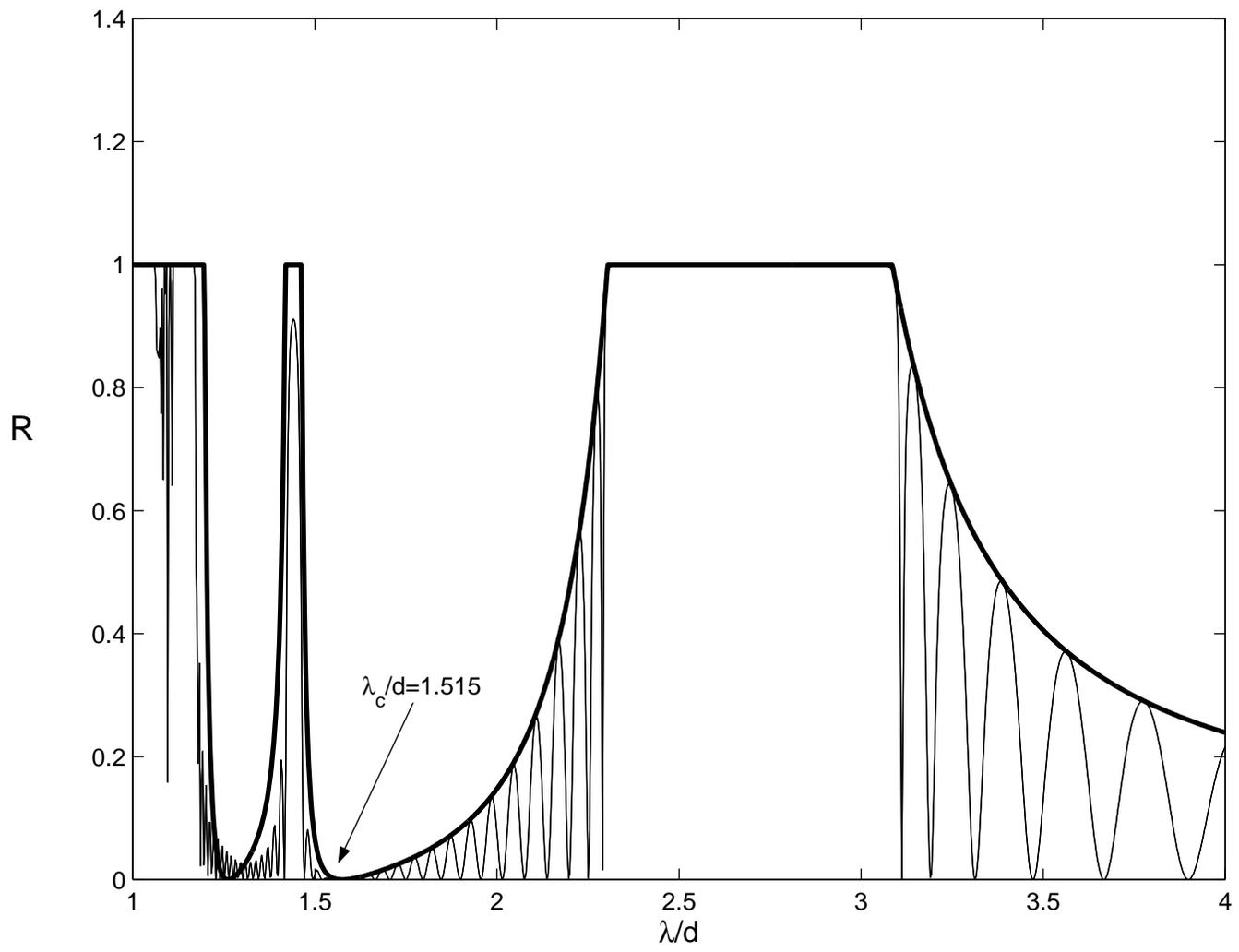
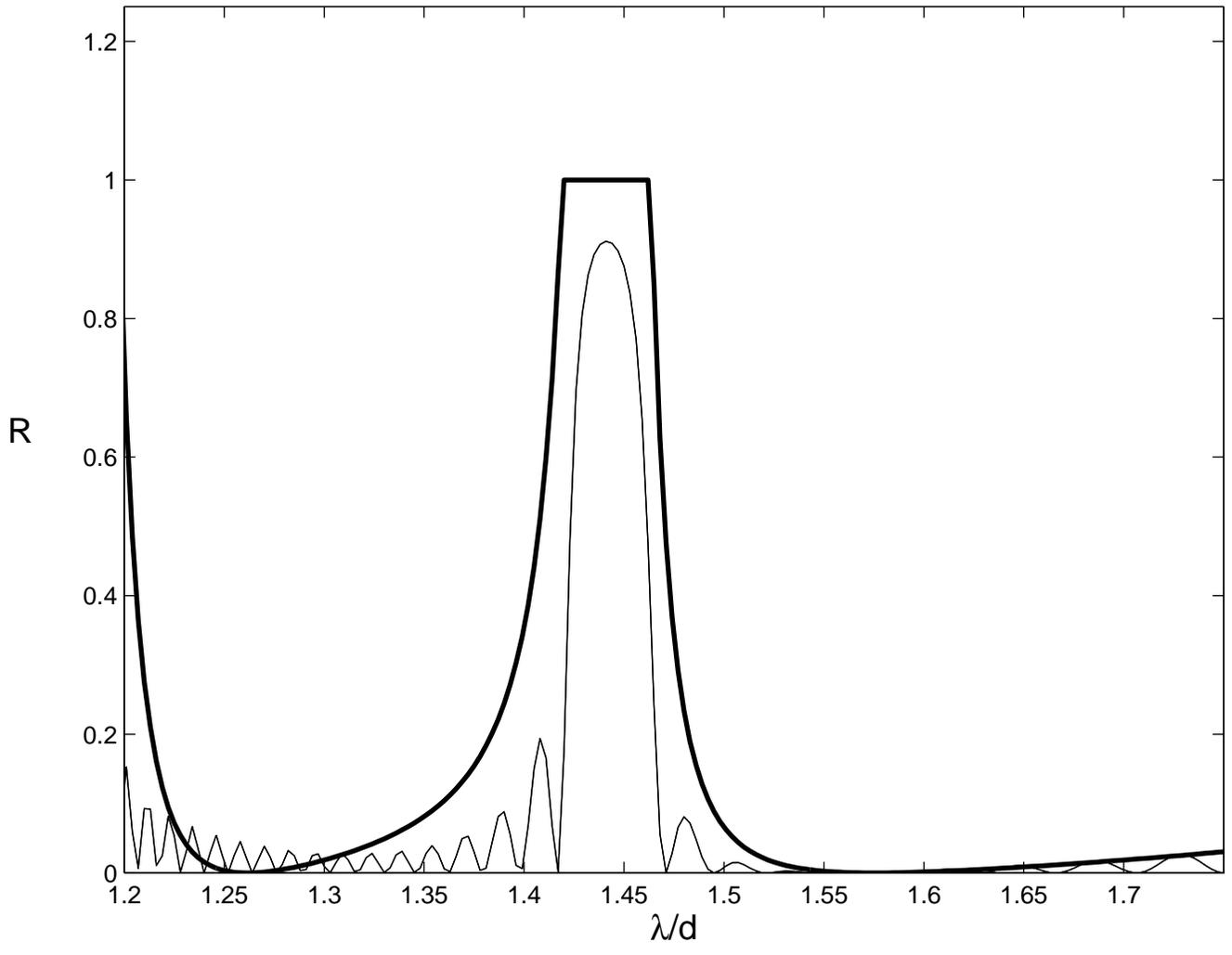
\end{document}